\documentclass[a4paper,11pt]{article}
\usepackage{palatino}
\usepackage{kotex}
\usepackage{authblk}
\usepackage{marvosym}
\usepackage{graphicx}
\usepackage{url}
\usepackage{apalike}
\usepackage{vmargin}
\setmarginsrb{2.5cm}{2.5cm}{2.5cm}{2.5cm}{0mm}{0mm}{0mm}{10mm}
\usepackage[colorlinks=true, allcolors=blue, breaklinks=true]{hyperref}

\title{\bfseries What is the planetary boundary layer?}
\author{\normalsize Natanael Karjanto\thanks{\Letter: \url{natanael@skku.edu} \href{https://orcid.org/0000-0002-6859-447X}{\includegraphics[scale=0.08]{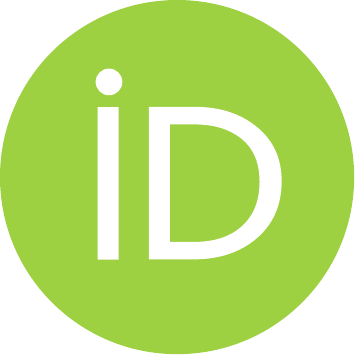}}}}
\affil{Department of Mathematics, University College, Natural Science Campus\\ Sungkyunkwan University, Suwon~16419, Republic of Korea}

\date{\vspace*{-0.5cm} \scriptsize Updated \today}

\begin{document}
\maketitle

\begin{abstract}
\noindent
This short article provides brief coverage of the planetary boundary layer (PBL) and some characteristics featuring this lowest part of the atmosphere. Any readers who are interested in deeper and more extensive coverage are encouraged to consult the references listed at the end of this article. \\

\end{abstract}

\section{Introduction}
This article covers a concise review of the planetary boundary layer (PBL). This part of the atmosphere should be of particular interest to many of us. Since we live within the PBL, our lives and those of vegetation and animals, flora and fauna, are directly influenced by what is happening here. From the moment of our first breath, we spend most of our lives near the Earth's surface. The majority of our activities take place in PBL, including growing plants, constructing houses, obtaining education, and working productively. We feel the warmth of the daytime sun and the chill of the nighttime air. We are accustomed to changes in local weather and conditions to long-term climate. When we travel to other parts of the Earth, we feel the difference in weather conditions. 

Recently, the Nobel Prize in Physics 2021 was awarded ``for groundbreaking contributions to our understanding of complex systems'' with one half jointly to Syukuro Manabe from Princeton University, US, and Klaus Hasselmann from the Max Planck Institute for Meteorology, Hamburg, Germany, ``for the physical modeling of Earth's climate, quantifying variability and reliably predicting global warming'' and the other half to Giorgio Parisi from Sapienza University of Rome, Italy, ``for the discovery of the interplay of disorder and fluctuations in physical systems from atomic to planetary scales.''~\cite{nobelprize2021} 

Interestingly, this year's physics prize is the first scientific Nobel awarded for understanding the climate. Furthermore, according to Thors Hans Hansson, chair of the Nobel Committee in Physics, this year's Laureates have all contributed to gaining deeper insight into the properties and evolution of chaotic, complex systems, and apparently random phenomena. The recognized discoveries demonstrate that our knowledge about the climate depends on a solid scientific foundation and is based on rigorous analysis of observations~\cite{press2021}.

In this brief coverage, we will discuss what PBL is and its characteristics. The readers who would like to delve deeper into the topic may conduct their research from the textbooks listed in the references. 

\section{Planetary Boundary Layer}

One definition of PBL is ``the part of the troposphere that is directly influenced by the presence of the Earth's surface and responds to surface forces with a timescale of about an hour or less.'' These forcings include not only transpiration and evaporation but also frictional drag, heat transfer, pollutant emission, and induced flow modification~\cite{stull1988introduction}. And although the definition of PBL includes the one-hour temporal scale, it does not mean that the PBL would reach equilibrium within that period, but rather, the transformation should have started in one hour or less.

\begin{figure}[h]
\begin{center}
\includegraphics*[width=0.45\textwidth]{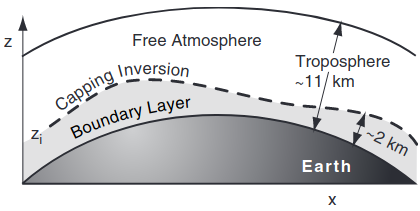}
\end{center}
\vspace*{-0.5cm}
\caption{A schematic diagram of the PBL shown as a vertical cross section of the Earth and troposphere, where the PBL occupies the lowest portion of the latter~\cite{stull2017meteorology}.}	\label{schema}
\end{figure}

PBL is also known by other names such as atmospheric boundary layer (ABL), mixed layer, boundary layer, or friction layer since this is the part of the atmospheric layer that is influenced by friction~\cite{ahrens2008essentials}. This friction occurs with the bottom boundary of the atmosphere, i.e., the Earth's surface. The thickness of the PBL may vary depending on space and time. It can range from tens of meters to more than 4~km, and the general thickness is around 1~or 2~km, i.e., filling the bottom 10\%--20\% of the troposphere~\cite{wallace2006atmospheric}. See Figure~\ref{schema}.
 
The PBL is characterized by transport processes, and turbulence is one of the essential ones, which is responsible for dispersing pollutants within the layer. Accumulation of pollutants in the PBL over several days may cause air pollution episodes in many urban regions. Through a particular mechanism, the turbulence both produces and is also influenced by heat flux, moisture, and momentum between the atmosphere and surface. The moisture flux moves upward through evaporation from the surface while the momentum flux via friction~\cite{randall2012introduction}. 

The lowest part of the atmosphere is known as the troposphere, and it includes the PBL. The part of the PBL that lies above the PBL is called the free troposphere. While the air in the PBL is turbulent, the free troposphere is characterized by positive static stability, i.e., buoyancy forces resist vertical motion. A turbulent process known as entrainment progressively assimilates the air in the free-troposphere into the PBL. While over the oceans the process is slow but steady, the entrainment over land is rapid and unstable. A strong daytime heating of the surface promotes turbulence and after the sunset, the entrainment diminishes and the PBL rearranges itself into a shallower nocturnal configuration~\cite{randall2012introduction}. 

While stable PBL occurs when the surface is colder than the air, such as during a clear night over land, or when warm air is advected over colder one, unstable PBL occurs when the opposite occurs, such as during a sunny day with light winds over the land, or when cold air is advected over a warmer water surface. With vigorous thermal updrafts and downdrafts, the latter is in a state of free convection. Neutral PBL forms during windy and overcast conditions and is in a state of forced convection~\cite{wallace2006atmospheric}.

\section{Conclusion}

In this short and brief exposition, we have considered the definition and characteristics of PBL. We hope to inspire the readers in seeking and investigating the topic further by reading the references below. Mathematical modeling and its corresponding physical interpretation certainly deserve further attention and should be addressed elsewhere.


\vspace*{1cm}
\subsection*{Dedication}
The author would like to dedicate this article to his late father Zakaria Karjanto (Khouw Kim Soey, 許金瑞) who introduced and taught him the alphabet, numbers, and the calendar in his early childhood. Karjanto senior was born in Tasikmalaya, West Java, Japanese-occupied Dutch~East~Indies on 1~January~1944 (Saturday~Pahing) and died in Bandung, West Java, Indonesia on 18~April~2021 (Sunday~Wage).

\begin{thebibliography}{99}
\bibitem[Nobel Prize, 2021]{nobelprize2021} The Nobel Prize in Physics 2021. NobelPrize.org. Nobel Prize Outreach AB 2021. Accessible online via \url{https://www.nobelprize.org/prizes/physics/2021/summary/}. Last accessed \today. 	

\bibitem[Nobel Prize Press Release, 2021]{press2021} The Nobel Prize in Physics 2021, Press Release (5 October 2021). Accessible online at \url{https://www.nobelprize.org/uploads/2021/10/press-physicsprize2021.pdf}. Last accessed \today.
	
\bibitem[Ahrens, 2008]{ahrens2008essentials} Ahrens, C. D. (2008) \textit{Essentials of Meteorology--An Invitation to the Atmosphere}, Fifth Edition. Belmont, CA: Thomson Brooks/Cole. 

\bibitem[Randall, 2012]{randall2012introduction} Randall, D. (2012) \textit{An Introduction to the Global Circulation of the Atmosphere}. Princeton, NJ: Princeton University Press. 

\bibitem[Stull, 1988]{stull1988introduction} Stull, R. B. (1988) \textit{An Introduction to Boundary Layer Meteorology}. Cham, Switzerland: Springer Science $+$ Business Media. 

\bibitem[Stull, 2017]{stull2017meteorology} Stull, R. B. (2017) \textit{Meteorology for Scientists and Engineers}, Third Edition. Vancouver, British Columnbia, Canada: University of British Columbia. 


\bibitem[Wallace \& Hobbs, 2006]{wallace2006atmospheric} Wallace, J. M. \& Hobbs, P. V. (2006) \textit{Atmospheric Science--An Introductory Survey}, Second Edition. Burlington, MA: Academic Press. 

\end{thebibliography}
\end{document}